\documentclass[12pt,preprint]{aastex}

\shorttitle{Strong Turbulence in Galaxy Clusters}
\shortauthors{Fujita et al.}

\begin{document}

\title{Strong Turbulence in the Cool Cores of Galaxy Clusters: \\ 
Can Tsunamis Solve the Cooling Flow Problem?}

\author{Yutaka Fujita\altaffilmark{1,2}, Tomoaki
Matsumoto\altaffilmark{3}, and Keiichi Wada\altaffilmark{1}}

\altaffiltext{1}{National Astronomical Observatory, Osawa 2-21-1,
Mitaka, Tokyo 181-8588, Japan; yfujita@th.nao.ac.jp}
\email{yfujita@th.nao.ac.jp}

\altaffiltext{2}{Department
of Astronomical Science, The Graduate University for Advanced Studies,
Osawa 2-21-1, Mitaka, Tokyo 181-8588, Japan}

\altaffiltext{3}{Department of Humanity and Environment, Hosei
University, Fujimi, Chiyoda-ku, Tokyo 102-8160, Japan}

\begin{abstract}
Based on high-resolution two-dimensional hydrodynamic simulations, we
show that the bulk gas motions in a cluster of galaxies, which are
naturally expected during the process of hierarchical structure
formation of the universe, have a serous impact on the core. We found
that the bulk gas motions represented by acoustic-gravity waves create
local but strong turbulence, which reproduces the complicated X-ray
structures recently observed in cluster cores. Moreover, if the wave
amplitude is large enough, they can suppress the radiative cooling of
the cores. Contrary to the previous studies, the heating is operated by
the turbulence, not weak shocks. The turbulence could be detected in
near-future space X-ray missions such as {\it ASTRO-E2}.
\end{abstract}


\section{Introduction}

Clusters of galaxies are the largest gravitationally bound and collapsed
systems in the universe. They are filled with X-ray emitted hot gas with
the temperature of $T\sim 2$--10~keV. High-resolution X-ray observations
have revealed that the hot gas in many cluster cores is not smoothly
distributed \citep[e.g.][]{fab01,san02}. Here, we define the cluster
cores as the dense regions within $\sim 30$--300~kpc from the cluster
centers. The complicated structures are often attributed to activities
of active galactic nuclei (AGNs). In fact, bubbles of high energy
particles have been found in some clusters
\citep[e.g.][]{fab00,mcn00,bla01}, which is the evidence that AGNs
affect the surrounding hot gas. However, the complicated structures have
also been observed in clusters in which AGNs are not active at the
centers \citep*[e.g.][]{fur03}. In the outer regions of the cores,
edge-shaped discontinuities in the gas density and temperature (`cold
fronts') are often found \citep*[e.g.][]{mar01}. They may be attributed
to the `sloshing' of the gas in the cluster gravitational potential
well, although the origin remains an open question \citep{mar01,chu03}.

From the above observations, one may think of complex gas motion, that
is, turbulence in the cores. Actually, turbulence is expected to be
prevailing in cluster cores. Although it has not yet directly been
observed in X-rays \citep[but see][]{chu04}, cold gas ($T\sim 10^4$~K)
moving with the velocity of 100--$1000\;\rm km\: s^{-1}$ has been
observed. If the ambient X-ray gas did not move with the cold gas, the
latter would immediately mix with the former \citep{loe90}. Moreover,
the turbulence may be playing an important role in heating of the
cluster cores. It had been expected that the gas in cluster cores cools
by radiating away its thermal energy, which induces gas motion toward
the cluster centers \citep[`cooling flows';][]{fab94}. However, X-ray
spectra have revealed that the mass of the cooling gas is far smaller
than that expected from this model \citep{mak01,pet01,kaa01,tam01}.  The
turbulence may transport thermal energy from the outside of the cluster
cores, and may balance the radiative cooling of the cores
\citep{cho03,kim03,voi04}. However, the actual mechanism that creates
the turbulence has not been understood.

The scenario we newly propose here is that the core turbulence is
created by bulk gas motions, which are naturally produced in the X-ray
hot gas in clusters. Cosmological numerical simulations have shown that
clusters are knots of larger-scale filaments in the universe
\citep{bur98}. Along the direction of the filaments, small clusters (or
galaxies) successively fall into a cluster, which increases the cluster
mass. The simulations have also shown that the infall velocities of the
smaller clusters are more than $1000\;\rm km\: s^{-1}$
\citep{bur98}. The interactions of the hot gas and dark matter between
the clusters should produce a large-scale gas motion; the velocity is
typically $\gtrsim 300\:\rm km\: s^{-1}$ and sometimes reaches $\gtrsim
1000\:\rm km\: s^{-1}$ \citep{bur98}. The cluster cores are exposed to
these violent gas motions; this would create possible turbulence and the
complicated structures observed in the cores even if there are no AGN
activities.

In order to test this scenario, we performed two-dimensional
high-resolution hydrodynamic simulations to follow the long-term
evolution of the core in `a stormy cluster' for the first time.

\section{Models}
\label{sec:models}

Since we are interested in the cluster core and we need resolution high
enough to resolve turbulence, we limited the calculations to the central
region of a cluster (within $\sim 300$~kpc from the cluster
center). These simulations were performed using a nested grid code
\citep{mat03}, and the coordinates are represented by $(R,z)$. While the
resolution near the outer boundary is $1.4$~kpc, that at the cluster
center is $22$~pc. Free boundaries were chosen. Thermal conduction,
viscosity, magnetic fields, and the self-gravity of gas were
ignored. Radiative cooling is included. We adopted a cooling function
for the metal abundance of 0.3 solar. The gas is isothermal and the
temperature is 7~keV at $t=0$. The fixed gravitational potential and the
initial gas distribution are the same as those in \citet*{fuj04b}. 

We approximated the bulk gas motions in a cluster by plane wave-like
velocity perturbations represented by $\delta v=\alpha c_s \sin(2\pi c_s
t/\lambda)$ at $z=-345$~kpc, where $c_s$ is the initial sound velocity,
and $\lambda$ is the wave length. This assumption comes from the idea
that continuous infall of matter and small clusters along a large-scale
filament should create velocity perturbations in a particular
direction. The factor $\alpha$ is a free parameter and $(R,z)=(0,0)$ is
the cluster center. We studied the perturbations with $\alpha$ and
$\lambda$ shown in Table~1. Note that cosmological numerical simulations
suggested that the velocity of the hot gas is about 0.2--$0.3\: c_s$ or
larger even when a cluster is relatively relaxed
\citep*{nag03,mot04}. Moreover, it is expected that the scale of bulk
gas motion in a cluster is $\sim 100$~kpc or larger
\citep*{roe99,mot04}. Therefore, we think that the parameters we took
are reasonable. However, since the large bulk motions in a cluster do
not always generate large amplitude acoustic-gravity waves, our model
may be appropriate for clusters undergoing hierarchical structure
formation or similar violent events. Since the energy input rate through
waves per unit area is given by $\sim (1/2)\rho \alpha^2 c_s^3$, where
$\rho$ is the gas density, the integrated input rate for $R<150$~kpc at
$z=-345$~kpc is $1.2\times 10^{44}(\alpha/0.3)^2\rm\: ergs\: s^{-1}$ for
our model cluster. On the other hand, at $t=0$, the X-ray luminosity of
the gas for $R<100$~kpc and $|z|<100$~kpc is $2.4\times 10^{44}\rm\;
ergs\; s^{-1}$ and is comparable to the wave energy input rate.

\section{Results}

\begin{figure}
\epsscale{.80} \plotone{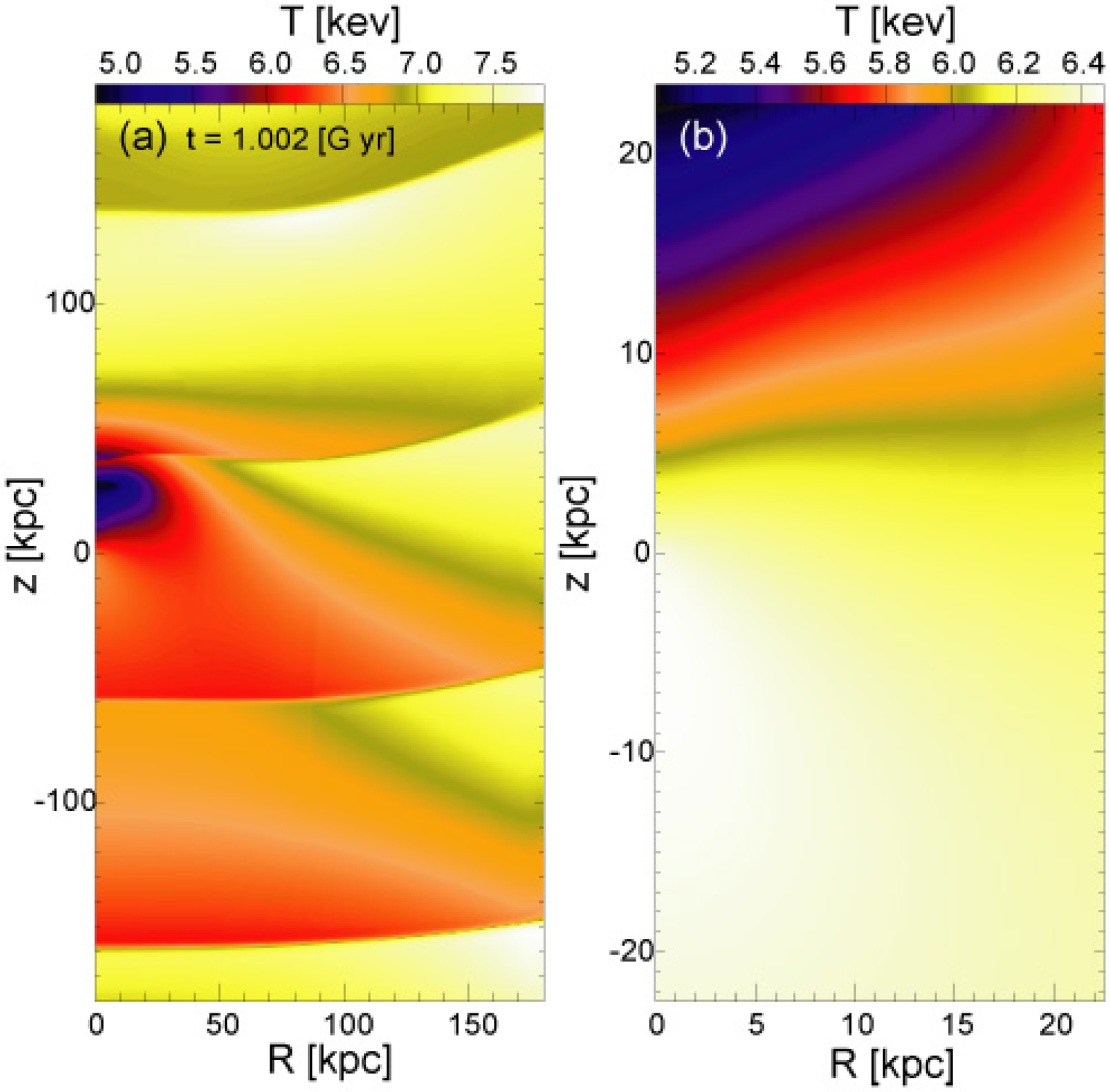} \caption{The temperature distribution of
a cluster core at $t=1.0$~Gyr. The periodic input waves are seen in
Fig.~\ref{fig1}a as discontinuities that is nearly parallel to the
$R$-axis. The waves propagate upward in these figures. (a) for
$z\lesssim 200$~kpc, (b) for $z\lesssim 20$~kpc. Movies are available at
http://th.nao.ac.jp/tsunami/index.htm. Note that
color bars are different between (a) and (b).  \label{fig1}}
\end{figure}

In Figures~\ref{fig1} and~\ref{fig2}, we present the temperature
distributions for $\alpha=0.3$ and $\lambda=100$~kpc; the velocity
perturbations propagate upwards as acoustic-gravity waves, which were
called `tsunamis' in \citet{fuj04b}. Since the velocity amplitude is
relatively large, the waves steepen and become weak shocks as shown in
one-dimensional simulations \citep{fuj04b}. Because of the pressure
coming from the momentum of the waves, the coolest and densest gas
noticeably shifts from the cluster center at $t\gtrsim 0.7$~Gyr and the
shift is clearly seen at $t=1$~Gyr (Fig.~\ref{fig1}). While the position
of the coolest gas is at $z\sim 25$~kpc (Fig.~\ref{fig1}), that of the
densest gas is $z\sim 10$~kpc. The difference of the positions is made
by the wave just passed (Fig.~\ref{fig1}); the gas slightly deviates
from pressure equilibrium there. Comparing the result of $\alpha=0.3$
with that of $\alpha=0$, we found that the gravitational energy of the
gas for the former is larger than that for the latter by $2.1\times
10^{60}$~ergs for $R<150$~kpc and $|z|<345$~kpc at $t=1$~Gyr. The energy
increase is smaller than the energy input by waves during the first one
Gyr for $R<150$~kpc ($3.9\times 10^{60}$~ergs), suggesting that a part
of the injected energy is used to produce the shift of dense gas. After
that, radiative cooling proceeds and the core region becomes cooler and
denser. Since waves no longer sustain the cooling core, the core falls
in the potential well of the cluster. During the fall, Rayleigh-Taylor
(RT) and Kelvin-Helmholtz (KH) instabilities develop around the
core. They non-linearly develop, and turbulent motion is eventually
formed in and around the core as seen in Figure~\ref{fig2}, which shows
the temperature distribution at $t=3.3$~Gyr. The cool core oscillates
around the cluster center, and temperature jumps are formed
(Fig.~\ref{fig2}a). Associated with these temperature jumps, gas density
also has discontinuities. Small cool and dense blobs randomly moving in
the core cause new RT and KH instabilities around them, and smaller
eddies are generated (Fig.~\ref{fig2}b). For the models of $\alpha=0.3$
and~0.5, the turbulence is maintained until our calculations are stopped
at $t\sim 5$~Gyr. It is interesting that the turbulence presented here
is naturally caused even by the regular, wave-like, linear
perturbations. No irregular triggers, which are usually essential for
initiating or maintaining turbulence, are necessary.

In Table~1, we present the time when the gas temperature in any of the
numerical grid points reaches zero ($t_{\rm cool}$). The gas cooling is
suppressed by heat transport through turbulent mixing, especially when
$\alpha$ is large. In other words, in addition to the wave input energy
(see \S\ref{sec:models}), the thermal energy transported from the
outside of the core is used to prevent the core from cooling.

We note that although there were several studies treating core heating
by acoustic-gravity waves, the mechanism of the heating presented here
is completely different from that proposed in the previous studies. In
those previous studies, based on the analytical approach \citep{pri89}
or one-dimensional numerical simulations \citep{fuj04b}, it was
predicted that weak shocks evolved from acoustic-gravity waves directly
heat the cluster core, while in this study the turbulence is the major
player of the heating. Since our simulations are multi-dimensional, the
results should be much more realistic than those of the previous
studies. In fact, since turbulence is essentially multi-dimensional,
those previous studies were not able to directly treat the turbulence
and complicated structures of the cores. Even with multi-dimensional
simulations, the turbulent heating could not be found if the resolution
were not high enough.

\section{Discussion}

The present results show that the turbulence starts to develop after the
core becomes dense through cooling. Before that, waves pass the cluster
center without much changing the gas structure. Thus, we predict that
this mechanism of turbulence generation works only for clusters with
dense and cool cores, which had been called `cooling flow
clusters'. Since the turbulence is spatially limited to the cool core,
it does not totally mix the hot gas in a cluster. Thus, as long as
violent mergers of clusters with comparable masses, which completely
destroy the central gas structures of the clusters, do not happen, the
metal abundance excess observed in cluster cores \citep[e.g.][]{tam01}
would not completely be erased.

\begin{figure}
\epsscale{.80}
\plotone{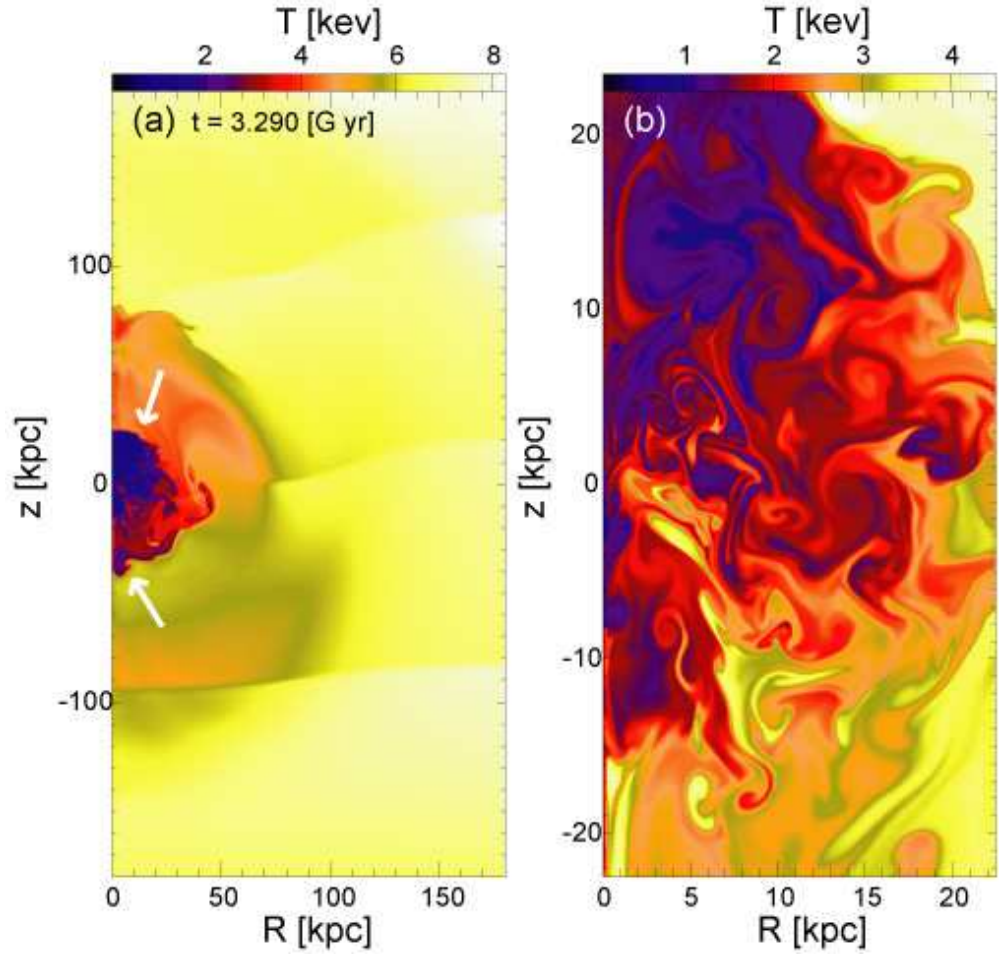}
\caption{The same as Fig.~\ref{fig1} but for $t=3.3$~Gyr. 
The arrows in Fig.~\ref{fig2}a indicate
 'cold fronts'. \label{fig2}}
\end{figure}

Because of the turbulent motion, the fine structures of the core are not
steady. Our simulation results sometimes show fine structures similar to
the peculiar structures observed in clusters such as A1795, Centaurus,
and 2A~$0335+096$ \citep*{fab01,san02,maz03}. The temperature jumps seen
in Figure~\ref{fig2}a may correspond to the `cold fronts' observed in
some clusters. The bulk gas motion in clusters would result in the
formation of acoustic-gravity waves and the weak shocks as shown
above. Direct observations of the waves may be difficult unless the wave
fronts are almost parallel to the line of sight. In A133, however, a
weak shock just passing through the core has been observed
\citep{fuj04a}.

The maximum velocity of the turbulence in a cluster core is $\gtrsim
300\:\rm km\: s^{-1}$ for $\alpha=0.3$. With a high spectral resolution
detector like the X-ray satellite {\it ASTRO-E2}, the turbulence in
cluster cores could be detected in the near future. If turbulence is
being developed, the metal lines in the X-ray spectra would have very
complicated features owing to the gas motion \citep{ino03}. On the other
hand, turbulence could also be created by AGN activities, especially by
the buoyant motion of AGN-origin bubbles. The lifetime of the eddies
associated with the bubble motion is $t_{\rm edd}\sim L_{\rm bub}/v_{\rm
bub}$, where $L_{\rm bub}$ and $v_{\rm bub}$ are the size and velocity
of a bubble, respectively. For the Virgo cluster, for example, the size
of the observed bubbles is $\sim 10$~kpc, and the predicted velocity of
a bubble is $\sim 400\:\rm km\: s^{-1}$ \citep{chu01}. Thus, the
lifetime of the eddies is $t_{\rm edd}\sim 2\times 10^7$~yr, which is
much shorter than the lifetime of the bubble itself \citep[$\sim
10^8$~yr;][]{chu01}. This means that turbulence is unlikely to exist in
a cluster core without AGN-origin bubbles. Thus, if turbulence is
detected in such a core, it could be associated with the bulk gas
motions outside of the core.

Our model predicts that cooling of a cluster core is more suppressed for
larger velocity amplitude. The suppression should also work in smaller
objects such as groups of galaxies and elliptical galaxies because their
overall structures are similar to those of clusters of galaxies
\citep{moo99}, and we expect that the bulk gas motions and turbulence are
excited in them by the same mechanism presented here. This is in
contrast with the suppression by thermal conduction that does not work
in the smaller objects because of their low temperatures \citep{voi04}.

Finally, we note the limitations of our model.  First, we approximated
the bulk gas motions by waves with constant amplitude and length, and
direction for propagation is also assumed to be fixed, but in reality
all these conditions should be changed during formation and evolution of
clusters. Even bulk gas motions that are not represented by regular
waves considered here could cause the motion of the cool cores and
produce turbulence through RH and KT instabilities. Moreover, it is
likely that waves come from several directions corresponding to
large-scale filaments. These effects should be ultimately studied by
fully three-dimensional, ultra-high-resolution cosmological simulations,
although the basic mechanism initiating the turbulence would not be
different from the one in the two-dimensional case here. Second, we did
not include the heating by turbulent dissipation, AGNs, and thermal
conduction. If they are effective as thermal energy sources, the wave
energy required to suppress the cooling of a core would be smaller than
that we predicted. Note that the turbulence may tangle magnetic fields
and reduce an effective conduction rate, However, \citet{nar01}
indicated that chaotic magnetic fields do not much reduce the conduction
rate. Thus, we may need to follow the evolution of magnetic field lines
when we include the effect of thermal conduction.  Third, we fixed the
gravitational potential of a cluster. A change in the potential may also
shift the cool gas core from the gravitational center of the cluster,
which may lead to the motion of the core and the development of
turbulence \citep{ric01}. Fourth, we assumed a two-dimensional
axi-symmetric geometry. This might affect development and structure of
turbulence.  It would be expected that cascade processes of eddies is
different in full three-dimensional turbulence. Interaction between the
turbulence and plane-waves in three-dimensions should be also clarified
in future simulations.

\acknowledgments

We thank the anonymous referee for useful suggestions. The authors were
supported in part by a Grant-in-Aid from the Ministry of Education,
Culture, Sports, Science, and Technology of Japan (Y. F.: 14740175;
T. M.: 14740134; K. W.: 15684003). All simulations were run on Fujitsu
VPP5000 at NAOJ.

{}

\clearpage

\begin{deluxetable}{ccc}
\tablecaption{Wave parameters}
\tablewidth{0pt}
\tablehead{
\colhead{$\alpha$} & $\lambda$ (kpc) & $t_{\rm cool}$ (Gyr)
}
\startdata
 0  & \nodata    & 2.2 \\
0.3 & 100 & 3.3 \\
0.5 & 500 & 4.7 \\
\enddata
\end{deluxetable}

\clearpage

\clearpage

\end{document}